\documentclass[aps,pra,showpacs,preprint]{revtex4}

\usepackage{bm}
\usepackage{amsmath}
\usepackage{amssymb}
\usepackage{graphicx}
\usepackage{epsfig}
\usepackage{epstopdf}
\begin{document}

\def\be{\begin{equation}}
\def\en#1{\label{#1}\end{equation}}
\def\d{\dagger}
\def\bar#1{\overline #1}
\newcommand{\per}{\mathrm{per}}

\newcommand{\rd}{\mathrm{d}}
\newcommand{\bx}{{\bf x}}
\newcommand{\bk}{{\bf k}}
\newcommand{\bq}{{\bf q}}
\newcommand{\br}{{\bf r}}
\newcommand{\bQ}{{\bf Q}}
\newcommand{\rX}{{\rm X}}
\newcommand{\vare}{\varepsilon }
\newcommand{\pd}{\partial}

\title{  Sufficient bound on the mode mismatch of single photons  for scalability of the   boson sampling computer     }

\author{V. S. Shchesnovich}

\address{Centro de Ci\^encias Naturais e Humanas, Universidade Federal do
ABC, Santo Andr\'e,  SP, 09210-170 Brazil }

\begin{abstract}

The boson sampler  proposed by  Aaronson and Arkhipov is  a non-universal quantum  computer, which   can serve as evidence against the extended Church-Turing thesis. It samples the  probability distribution at the output of linear  unitary optical  network, with indistinguishable single photons at the input. Four experimental  groups have already tested their small-scale prototypes with up to four photons.  The boson sampler  with few dozens  of   single photons is believed to  be  hard to  simulate on  a classical  computer.  For scalability of a realistic boson sampler with  current technology it is necessary to know the effect of the photon  mode mismatch on its operation.   Here a nondeterministic model of the boson sampler  is analyzed, which employs   partially indistinguishable single photons emitted by identical sources. A sufficient condition on the average mutual fidelity $ \langle \mathcal{F}\rangle$  of the single photons is found,  which guarantees that the  realistic boson sampler outperforms the classical computer.   Moreover, the  boson sampler computer with partially indistinguishable single  photons is scalable   while being beyond the power of classical computers when  the single photon  mode mismatch  $1-\langle \mathcal{F}\rangle$  scales as $ \mathcal{O}(N^{-3/2})$ with the total  number of  photons  $N$.
\end{abstract}

\maketitle

\section{Introduction}

The boson sampler (BS) computer proposed recently by  Aaronson and Arkhipov \cite{AA} can serve as   evidence  against the  extended Church-Turing (ECT) thesis which says that any physical device can be efficiently simulated on the probabilistic Turing machine. No interaction between  bosons is required, thus the BS computer can be built using  only passive  linear optical devices and  emitters of    indistinguishable  single photons \cite{SPS}, i.e. the single photons  producing  the Hong-Ou-Mandel type interference \cite{HOM} (see, also Refs. \cite{LB,MPI}).  Whereas the universal quantum computer    targets the  $NP$ decision problems,  widely believed to be classically hard, such as factoring large integers  \cite{Shor,BookNC}, the BS computer  just samples  the    output probability  distribution of  $M$-mode  unitary   network $U$ with $N$ identical bosons    at its input.   It is shown that simulation  of the BS   on a classical computer  requires   exponential resources in the number of bosons $N$ (when $M\ge N$) \cite{AA}, since    bosonic    amplitudes    are given as the  permanents  (see  Ref. \cite{Minc} for the  definition and properties) of complex $N\times N$-submatrices of $U$  \cite{C,S}, whose    computation  is exponentially hard \cite{Valiant,A1}  (the fastest known Ryser's  algorithm \cite{Ryser} requires $\mathcal{O}(N^22^N)$ flops).   On the conceptual side,    a classical algorithm for the matrix permanent would provide also for solution of all problems in the complexity class $\#P$, of a higher complexity than the $NP$ class, which, in its turn, would imply  dramatic theoretical consequences: collapse of the whole polynomial hierarchy of the  computational complexity  \cite{AA}. While an universal quantum computer can simulate the BS, the scalability of the BS beyond the classical computational power   is easier to achieve: already with   $  20 \le N \le 30$ photons it would outperform  the classical computers \cite{AA}.     Four independent groups have  already tested their prototypes of the BS on small networks with up to four input  photons   \cite{E1,E2,E3,E4}.  

It is crucial that    even  an \textit{approximate} simulation of the BS computer  must  be classically hard (at least  when  $M\gg N^2$) \cite{AA}, hence,   the stringent  fault-tolerances required for the universal quantum computer \cite{ErrCorr,ErrCorrRev,Steane,Knill}  may be  significantly relaxed for the BS computer.    The necessary, though not sufficient, conditions  for the BS operation beyond the power of classical computers were analyzed in Refs. \cite{RR,R1}, supporting this view. It was even suggested \cite{R1} that scaling up helps to combat photon  mode mismatch and losses.   Recently, the effect of noise in the experimental realization of a unitary network on the BS complexity was  studied \cite{NoisyBS}. It was shown that  fidelity of the  optical elements must be at least  $1- \mathcal{O}(N^{-2})$ for the noisy-network  realization of the  BS to be still hard to simulate classically.   These results suggest the experimental feasibility of the BS computer in the near future.

In  practice,  limitations on  indistinguishability of single  photons from  realistic sources will be always present.   All four groups of Refs.~\cite{E1,E2,E3,E4}  have  tested their BS prototypes  using  the so-called  heralded single photons from the  parametric down conversion,  not free from  the  multi-photon  components and noise.    It is clear that  some amount  of  indistinguishability of  single photons is   essential for the BS computer  (a large  mode mismatch  allows for an efficient simulation on a classical computer   \cite{AA} by a probabilistic  algorithm \cite{JSV}, see also below).
 Recently a spatial multiplexing of the heralded single photon sources was proposed to  enhance the relative  yield of the  single photon component   \cite{Multiplex}, but   scalability is still out of reach.  On the other hand,  scalable  single photon sources with  high photon  antibunching   can be  based on individual emitters such as the quantum dots     \cite{NDSPS1,NDSPS2,NDSPS3}, but they are inherently nondeterministic, since based on the  spontaneous emission or on the spontaneous decay from a cavity.   If the  nondeterministic sources of single photons could be employed to scale up the BS? Generally, what specific features of bosonic particles are necessary for  the BS computer to outperform the classical computer?   A related fundamental problem is that, to date,  no \textit{sufficient } bound  is known  on the  mode mismatch  of  single photons for  experimentally realistic BS to serve as   evidence against the ECT. 

\textit{Thus,   it is of paramount importance  for  building a scalable BS device to establish   the  degree of distinguishability of  single photons   for the  BS to be still hard to simulate on a  classical computer. }  This is the main focus of the present work. The analysis is concentrated  on the effect of the  photon   mode mismatch by neglecting  two other sources of error, i.e.  noise in experimental realization of  an unitary network and photon losses. A sufficient bound on the mode mismatch is derived for the BS computer with partially indistinguishable single photons to outperform the classical computer. \textit{For instance, the  BS computer with partially indistinguishable single  photons  is scalable  beyond the power of the classical computer if the  mode mismatch  $1-\langle \mathcal{F}\rangle$, where  $ \langle \mathcal{F}\rangle$ is the average single photon fidelity, scales as $ \mathcal{O}(N^{-3/2})$ with the total  number of  photons  $N$.} In derivation of the fidelity bound,  the indistinguishability of $N$ single photons in distinct  modes  is quantified by a $N$-vector parameter -- an  approach which  can be useful  in other problems. 

The rest of the text is organized as follows. In section \ref{sec2} the nondeterministic boson sampler (NDBS) model  is formulated, which captures the essential features of any non-ideal  BS computer  with  the  single  photons only partially indistinguishable. Section  \ref{sec3}  is devoted to analyzing the conditions under which the NDBS performs a classically hard computational task. In section \ref{sec4} a short summary of the results is given. Some inessential   mathematical  details of the derivations and other computational  details are relegated to Appendices \ref{secA1},  \ref{secA3}, and  \ref{secA4}.

\section{The Nondeterministic Boson Sampler  model}
\label{sec2}

\begin{figure}[htb]
\begin{center}
\includegraphics[width=0.35\textwidth]{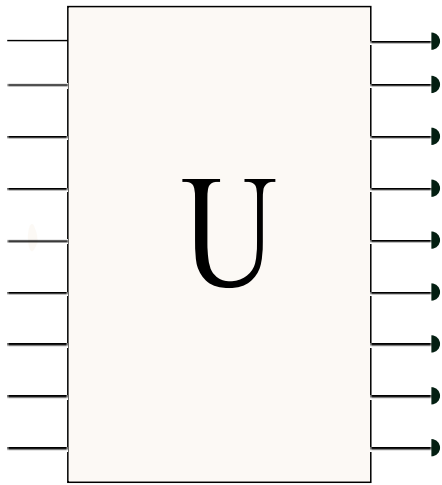}
\caption{  Schematic (black-box) depiction of the  NDBS setup with the network matrix $U$, where on the left are the input modes corresponding to the operators $a_k(\omega)$, linked to the identical photon sources, and on the right are the output modes corresponding to the operators $b_k(\omega)$ linked to  the detectors.    }\label{Fig1}
\end{center}
\end{figure}

Consider   $N$  single  photons emitted by identical    sources and  launched  into distinct input  modes $k_1, ..., k_N$   of a $M$-mode   linear optical  network given by an unitary matrix $U$: $a^\dag_k (\omega)= \sum_{l=1}^M U_{kl}b^\dag_l(\omega)$, where $a_k(\omega)$ and $b_k(\omega)$  are  the  input  and output  modes of frequency $\omega$, respectively (see Fig. \ref{Fig1}).
The input state is given by a density matrix. Setting $\mathbf{x}$ to be a fluctuating  vector-parameter  in the spectral function $\phi(\mathbf{x},\omega)$ of a  single  photon (for instance, the arrival time or phase) with the distribution  $p(\mathbf{x})$, identical for each source, the density matrix reads $\rho^{(in)} = \int\rd \mathbf{x}_1\!...\!\int\rd\mathbf{x}_N\left[ \prod_{\alpha=1}^Np(\mathbf{x}_\alpha)\right]|\Psi(\mathbf{x}_1,...,\mathbf{x}_N)\rangle \langle \Psi(\mathbf{x}_1,...,\mathbf{x}_N)|$, where
\be
|\Psi(\mathbf{x}_1,...,\mathbf{x}_N)\rangle  = \prod_{\alpha=1}^N\int\limits_0^\infty\rd \omega_\alpha \phi(\mathbf{x}_\alpha,\omega_\alpha)a^\dag_{k_\alpha}(\omega_\alpha)|0\rangle
\en{EQ1}
is a Fock state of $N$ photons at the input. This is a more general setup than in Ref.~\cite{AA}, which allows to consider the effect of photon mode mismatch.   The output probability  of detecting $m_1,...,m_M$ photons in modes $1,...,M$ can be derived  by the  quantum photon counting  theory \cite{Mandel,Glauber,KK}.  The result is that the     probability  is  given by   the following positive  Hermitian  operator  (see Appendix \ref{secA1})
\begin{eqnarray}
\label{Pi}
&&\Pi(m_1,...,m_M) = \frac{1}{\prod\limits_{l=1}^M m_l!}\int\limits_0^{\infty} \rd \omega_1...\int\limits_0^{\infty} \rd \omega_N\prod_{\alpha=1}^N\Gamma(\omega_\alpha)\nonumber\\
&&\times \left[ \prod_{\alpha=1}^N b^\dag_{l_\alpha}(\omega_\alpha)\right]|0\rangle\langle 0|\left[\prod_{\alpha=1}^N b_{l_\alpha}(\omega_\alpha)\right],
\end{eqnarray}
where $(l_1,...,l_N)\equiv \{1,...,1,2,...,2,...,M,...,M\}$,  with index $j$ appearing $m_j$ times, and
$\Gamma(\omega)\ge0$ is the spectral function  of the  detector. The set of all such   operators  as in   Eq. (\ref{Pi}), after a suitable normalization (see below),   constitute the POVM describing  photon detection at the  output modes.  By Eqs. (\ref{EQ1}) and (\ref{Pi}), the detection probability $P(m_1,...,m_M|k_1,...,k_N) = \mathrm{tr}\{\Pi(m_1,...,m_M)\rho^{(in)}\}$  becomes
\begin{eqnarray}
\label{EQ2}
&& \!\! P(m_1,...,m_M|k_1,...,k_N) = \frac{1}{\prod_{l=1}^M m_l!}\sum_{\sigma_1}\sum_{\sigma_2} J(\sigma_2\sigma^{-1}_1) \nonumber\\
&& \quad \times \prod_{\alpha=1}^N U^*_{k_{\sigma_1(\alpha)},l_\alpha} U_{k_{\sigma_2(\alpha)},l_\alpha}
\end{eqnarray}
with each   sum running  over all permutations of $N$ indices $k_1,...,k_N$  in  a $N\times N$-submatrix of the network matrix $U$.
In fact,  Eq. (\ref{EQ2}) applies  more generally, not necessarily with  identical sources,   when the network  input consists of  states with up to one photon per mode.  In this  general case,   $J$  depends only  on the relative permutation $\sigma_{21}\equiv\sigma_2\sigma^{-1}_1$~\footnote{Since $J$ is independent of $U$,   the  substitution  $\alpha = \sigma^{-1}_1(\beta)$ gives  $U^*_{k_{\beta},l^\prime_\beta} U_{k_{\sigma_{21}(\beta)},l^\prime_\beta}$, i.e. the transformed   matrix $U$ with the $\sigma_1$-permuted output indices  $\l^\prime_{\beta} \equiv l_{\sigma^{-1}_1(\beta)}$. Hence $J$ depends on $\sigma_{21}$ only.}.  Evidently $J = \delta_{\sigma_1,\sigma_2}$  is the classical limit, whereas the ideal BS of Aaronson and Arkhipov  has  $J = 1$ (independently of its argument).  In our case, due to  identical  sources, $J$   factorizes into a product of functions of  cycles   of the relative permutation, where  cycles of the same length contribute the same factor~\footnote{Since the sources are  identical, two cycles involving  $k$ sources  can be mapped into each other by relabeling the sources.}. Thus  $J$ is a function  of the cycle structure  $C_1,...,C_N$ of $\sigma_{21}$,  ($C_k$ is the number of cycles of length $k$,  $\sum kC_k=N$ \cite{Stanley}). In particular, we obtain (see Appendix \ref{secA1})
\be
J(\sigma)  = \prod_{k=2}^N g_k^{C_k(\sigma)},
\en{EQ3}
where we have introduced
\be
g_k=\prod_{\alpha=1}^k\int\rd \mathbf{x}_\alpha p(\mathbf{x}_\alpha)\int\limits_0^\infty\rd \omega_\alpha
\Phi(\mathbf{x}_{\alpha},\omega_{\alpha-1})\Phi^*(\mathbf{x}_\alpha,\omega_\alpha)
\en{gk}
with  $\Phi(\mathbf{x},\omega) \equiv \sqrt{\Gamma(\omega)}\phi(\mathbf{x},\omega)$ (the product is a shortcut notation   for the multiple integrals over
 $\mathbf{x}_\alpha$ and $\omega_\alpha$, where $\alpha=0$ is the same as  $\alpha=k$).  For   efficient broad-band detectors   a small percent of losses can be dealt with  the postselection. In this case, normalizing   the modified spectral function as $\int\rd \omega |\Phi(\mathbf{x},\omega)|^2 = 1$, we get for the probabilities of Eq.~(\ref{EQ2}): $\sum_{\{m_j\}}P(m_1,...,m_M|k_1,...,k_N) =   1$, where the summation  is constrained by $m_1+...+m_M=N$ (indeed, the described  renormalization  is equivalent to setting $\Gamma(\omega) = 1$, i.e. to the case of  bandwidth unlimited ideal detectors and    single photons with the modified spectral function, where all photons are detected).  

The   $g_k$  has   physical meaning of $k$-photon indistinguishability  parameter defined for identical single photon sources  (in general,     indistinguishability of single photons is described by the Young diagrams  \cite{Young};  for general multi-photon case see Ref.~\cite{Ou,RMS}).  In the ideal BS case all $g_k=1$, whereas the classical case is $g_k=0$, $k\ge 2$. The physical meaning of $g_k$ requires that it is positive. This and other properties of $g_k$ can be easily seen from the following representation. Introduce the following one-particle  density matrix 
\be
{\rho} \equiv \int\rd \mathbf{x}\, p(\mathbf{x})|\Phi(\mathbf{x})\rangle\langle \Phi(\mathbf{x})| 
\en{rho}
with vector $|\Phi(\mathbf{x})\rangle \in \mathcal{H}$ defined as   $\langle \omega|\Phi(\mathbf{x})\rangle \equiv \Phi(\mathbf{x},\omega)$, where the  Hilbert space $\mathcal{H}$  has the resolution of unity given by   $\int\limits_0^\infty\rd \omega |\omega\rangle\langle \omega| = \hat{1}$.  Note that  the above normalization of $\Phi(\mathbf{x},\omega)$ guarantees that $\mathrm{tr}(\rho)=1$. 
Under these definitions,   Eq. (\ref{gk}) can be cast in the  form  of a trace of a positive operator (by recognizing in the  integrals the above defined resolution of unity in $\mathcal{H}$) 
\begin{eqnarray}
\label{E13}
&&g_n = \int\limits_0^\infty\rd \omega_1\ldots\int\limits_0^\infty\rd \omega_n \prod_{j=1}^n \langle\omega_j|\rho|\omega_{j+1}\rangle\nonumber\\
&& = \int\limits_0^\infty\rd \omega_1 \langle\omega_1|\rho^n|\omega_1\rangle = \mathrm{tr} \rho^n.
\end{eqnarray}
Hence,  $0\le g_n\le 1$.  Moreover, passing in  the diagonal basis, we also obtain an important bound for  higher  indistinguishability parameters (setting also $g_1=1$, for convenience) 
\begin{equation}
\label{E14}
g_n  = \mathrm{tr}(\rho^k \rho^{n-k}   )
\le\mathrm{tr}(\rho^k)  \mathrm{tr}(\rho^{n-k} )= g_kg_{n-k}. 
\end{equation}
For instance, $g_{n+1}\le g_{n}$.

One general observation follows: since the  computational complexity of the NDBS  decreases as  $J(\sigma)$ deviates from its maximum $J=1$~\footnote{As follows  from Eq. (\ref{EQ2}), such a deviation   inserts small coefficients at the  products  of quantum amplitudes $\prod_{\alpha=1}^N U^*_{k_\alpha,l_\alpha}U_{n_\alpha,l_\alpha}$  of the ideal BS, thus reducing the  number of effectively contributing  terms (for fixed $k_1,...,k_N$), which varies from  $1$ ($n_\alpha=k_\alpha$, the classical case) to $N!$  (the ideal BS case).} (except on the identity permutation) and the   indistinguishability parameters satisfy   $g_{n+1}\le g_n$, it  is  doubtful   that   scaling up to higher number of single photons can help to combat the photon mode mismatch (as suggested  in  Ref. \cite{R1}).   Below we derive a sufficient  condition on the mode mismatch which  has an inverse $3/2$-power law scaling in the total  number of photons.

Eqs. (\ref{EQ2})-(\ref{gk}) are the basis of our consideration.    Below we focus on the region  of  small mode   mismatch.  In this case the  average mutual fidelity  of the single photons  (denoting the averaging over $\mathbf{x}$  by $\langle\ldots\rangle$)
\begin{eqnarray}
&&\langle\mathcal{F}\rangle =\int\rd\mathbf{x}_1 p(\mathbf{x}_1)\int\rd \mathbf{x}_2p(\mathbf{x}_2)|\langle \Phi(\mathbf{x}_1)|\Phi(\mathbf{x}_2)\rangle|\nonumber\\
&&= \int\rd\mathbf{x}_1 p(\mathbf{x}_1)\int\rd \mathbf{x}_2p(\mathbf{x}_2)\left|\int\rd\omega
\Phi^*(\mathbf{x}_1,\omega)\Phi(\mathbf{x}_2,\omega)\right|\nonumber\\
&&\label{F}\end{eqnarray}
 can be expanded in powers of the vector variable $\mathbf{x}$ (we set, for simplicity,  
 $\langle \mathbf{x}\rangle = \mathbf{0}$). Indeed from Eq. (\ref{F}), using that $\mathbf{x}_{1,2}$ have identical distributions, we get  
  \begin{eqnarray}
&&\langle\mathcal{F}\rangle = \biggl\langle \biggl[\int\rd\omega
\Phi^*(\mathbf{x}_1,\omega)\Phi(\mathbf{x}_2,\omega)\nonumber\\
&&\times\int\rd\omega^\prime
\Phi(\mathbf{x}_1,\omega^\prime)\Phi^*(\mathbf{x}_2,\omega^\prime)\biggr]^\frac{1}{2}\biggr\rangle \nonumber\\
&&=1 - \sum_{i,j}A_{ij}\langle x_i x_j\rangle +\mathcal{O}(\langle\mathbf{x}^3\rangle),
\label{F1}\end{eqnarray}
 where  we have used that $\mathbf{x}$ is real and defined a symmetric  (necessarily positive) matrix given  by the  photon sources:
\begin{eqnarray}
&&A_{ij} = -\mathrm{Re}\biggl\{ \int\rd\omega \Phi^*(\mathbf{0},\omega)\frac{\partial^2\Phi(\mathbf{0},\omega)}{\partial{x_i}\partial {x_j}} \nonumber\\
&&+   \int\rd\omega \Phi(\mathbf{0},\omega) \frac{\partial\Phi^*(\mathbf{0},\omega)}{\partial{x_i}} \int\rd\omega^\prime\Phi^*(\mathbf{0},\omega^\prime)\frac{\partial\Phi(\mathbf{0},\omega^\prime)}{\partial{x_j}}\biggr\}.\nonumber\\
&&\label{Aij}\end{eqnarray}
 One important relation can be also established between $g_k$ and $ \langle\mathcal{F}\rangle$ for small mode mismatch. Indeed,   the  single-particle density matrix (\ref{rho}) has the following expansion  in power series of $\mathbf{x}$
\be
\rho =  |\Phi(\mathbf{0})\rangle\langle \Phi(\mathbf{0})|  - \sum_{ij}\mathcal{A}_{ij}\langle x_ix_j\rangle  + \mathcal{O}(\langle\mathbf{x}^3\rangle),
\en{Aprho}
where the operator $\mathcal{A}_{ij}$ reads
\begin{eqnarray}
\label{Aop}
\mathcal{A}_{ij} &=& -\frac{1}{2} \left[ |\Phi(\mathbf{0})\rangle\langle \frac{\partial^2\Phi(\mathbf{0})}{\partial {x}_i\partial {x}_j}| + |\frac{\partial^2\Phi(\mathbf{0})}{\partial {x}_i\partial {x}_j}\rangle\langle \Phi(\mathbf{0})\right]\nonumber\\
&&-|\frac{\partial\Phi(\mathbf{0})}{\partial x_i}\rangle\langle \frac{\partial\Phi(\mathbf{0})}{\partial x_j}|.   
\end{eqnarray}
Then, utilizing Eq. (\ref{E13}), noticing that $\mathrm{Re}\left(\langle\Phi(\mathbf{0})|\mathcal{A}_{ij}|\Phi(\mathbf{0})\rangle \right)=  A_{ij}$ defined in Eq. (\ref{Aij}),  and comparing with  Eq. (\ref{F1}) the following important relation is  established: $g_k = 1 - k(1 - \langle\mathcal{F}\rangle)+\mathcal{O}(\langle\mathbf{x}^3\rangle)$, i.e. for a small mode  mismatch,  the $k$-photon  distinguishability  parameter $1-g_k$ is $k$ times the mode mismatch (defined here as the deviation of the average   fidelity $\langle \mathcal{F}\rangle$ of Eq.  (\ref{F})  from $1$).

One important model, in view of nondeterministic sources,  is  of the photons with random arrival times $\tau$ (equivalently, random  phases), where $\Phi(\tau,\omega) =    \phi(\omega)e^{i\omega \tau}$ (we set $\langle\tau\rangle=0$).  Let us denote the standard deviation (i.e. dispersion) of the arrival times by $\Delta\tau$,  that of the frequency by $\Delta \omega$ (under  the spectral density $|\phi(\omega)|^2$), and introduce  the classicality parameter   $\eta = \Delta\omega\Delta\tau$ (for $\eta = 0$ we recover the BS of Aaronson and Arkhipov, while for $\eta = \infty$ the classical case). Then we obtain 
 $\langle \mathcal{F} \rangle= 1 - \eta^2 + \mathcal{O}(\eta^4)$. Similarly, we also have $g_k(\eta) = 1 - k\eta^2 +\mathcal{O}(k^2\eta^4)$ giving $J(\sigma)= 1 - [N-C_1(\sigma)]\eta^2 + \mathcal{O}(N^2\eta^4)$. These expressions for a  small mismatch follow also from the general case, where one can identify $\eta^2 =  \sum_{i,j}A_{ij}\langle x_ix_j\rangle$ and $A_{ij}$ defined in Eq. (\ref{Aij}) (however,  generally, the  order of the next term is $\mathcal{O}(\langle\mathbf{x}^3\rangle)$, whereas the absence of the third-order term for the random arrival times model is due to a single fluctuating  parameter $\tau$ and  the fact that  $\langle \mathcal{F}\rangle$ and $g_k$ are symmetric w.r.t. permutations of  the integration variables $\tau_i$ and only their differences $\tau_i-\tau_j$ enter the definitions). Thus one can think of $[\sum_{i,j}A_{ij}\langle x_ix_j\rangle]^{1/2}$ as an  analog of the  classicality parameter in the general case (at least for a small mode mismatch).

\section{The Nondeterministic Boson Sampler and a  classically hard computational task}
\label{sec3}

The hardness result  of  Aaronson and Arkhipov  \cite{AA}  is formulated for the Haar-random  network matrix $U$ in the dilute limit (defined here as $M\gg N^2$), assuring that the submatrices of such a random matrix  are approximated by  matrices with the elements being i.i.d. Gaussians with $\langle U_{kl}\rangle = 0$ and
$\langle |U_{kl}|^2\rangle = \frac{1}{M}$ (since $\sum_{l=1}^M|U_{kl}|^2 =1$). The  distribution density of  elements  of $U$ factorizes in this approximation and is given by~\footnote{It is proven   that for $M\ge (N^5/\epsilon)\mathrm{log}^2(N/\epsilon)$  the Haar  probability density $p_H$ satisfies  $p_H(X)\le (1+\mathcal{O}(\epsilon))p(X)$, where $p$ is the probability density of Eq. (\ref{EQ4}), but   a similar relation  is  expected to be valid  for $M\gg N^2$ \cite{AA}.}
\be
p(U_{kl}) = \frac{M}{\pi}\exp\{-M |U_{kl}|^2\}.
\en{EQ4}
The dilute  limit  is also essential for practical implementation, since one can use the simplest on-off  (a.k.a. bucket)  photon  detectors, because of the  vanishing probability of multi-photon detection at the output  modes,  due to  the  ``boson birthday paradox"  \cite{AA,AG}, now experimentally verified  \cite{BBPExp}, which is similar to the classical birthday paradox.  Therefore, we  can  restrict ourselves to  the output occupation numbers $m_l \in \{0,1\}$, introducing     $l_1,...,l_N$ as the   \textit{distinct}  output modes (denoting $\vec{l} \equiv (l_1,\ldots,l_N)$, etc)  and  setting $P_\eta(\vec{l}|\vec{k})$ to be the corresponding   output probability. Note that the  sum of  probabilities of the bunched outputs is small on average over the Haar measure, being  on the order of $\mathcal{O}( N^2/M)$ \cite{AA}. 

The main result of  Aaronson and Arkhipov \cite{AA}  states that   approximation of the ideal BS   cannot be performed on a classical computer with only polynomial resources in  the total number of photons $N$ and  inverse of  the approximation error. The   approximation error $\varepsilon$  is  the variational distance of the output distributions between the ideal BS case, $\mathcal{D}_0$, and the proposed approximation, $\mathcal{D}_1$.  In our case,  the above means that  the NDBS is classically hard to simulate in polynomial time in $(N,1/\varepsilon)$   if, for a Haar-random  network matrix $U$, its output distribution $\mathcal{D}_\eta$  on the single photon outputs is variationally close to that of the ideal BS, i.e.
\be
||\mathcal{D}_0-\mathcal{D}_\eta||^\prime \equiv \frac{1}{2} \sum_{\vec{l}}|P_0(\vec{l}|\vec{k}) - P_\eta(\vec{l}|\vec{k})  | \le c\varepsilon,
\en{VarDist}
for some fixed constant $c$. Indeed,  the (average in the Haar measure)  probability to have a  bunched output is vanishing as  $\mathcal{O}( N^2/M)$, thus  the correction to  the variational distance, i.e. the  difference   between the complete  and the nonbunched outputs, satisfies (on average) $ ||\mathcal{D}_0-\mathcal{D}_\eta|| - ||\mathcal{D}_0-\mathcal{D}_\eta||^\prime =\mathcal{O}( N^2/M) \ll 1 $.

The main point of the arguments in Ref. \cite{AA} is that an approximation of the BS computer as above described also  solves  some   computational task impossible to solve on a classical computer. Specifically, it   was shown that such a classical simulation  would  imply also  approximation of  the permanents of matrices  of   Gaussian i.i.d. complex random variables with only polynomial resources, which is conjectured   to be impossible (some numerical and other evidence is provided).   Below, we will use one of the equivalent formulations of the latter computational task,  namely,  the problem to approximate the probability of the ideal BS to within an additive error  $\pm \varepsilon \langle P_0(\vec{l}|\vec{k})\rangle = \pm \varepsilon \frac{N!}{M^N}$, where the average with  respect to the Haar measure is computed using  the Gaussian approximation (\ref{EQ4})  (under the Gaussian approximation, this problem is equivalent to $|GPE|^2_\pm$ of Ref. \cite{AA}). Let us formulate it in precise terms.

\textit{$|BS|^2_\pm$-problem.}  For the  \textit{ideal}  BS computer  with a Haar-random  $M\times M$-dimensional unitary network  matrix $U$ and $N$ single photons at the input,  given  small parameters $\varepsilon$ and $\delta$, simulate  the output probability  $P_0(\vec{l}|\vec{k})$  to within the additive  error $\pm \varepsilon \frac{N!}{M^N}$, with  success probability (in the Haar  measure) at least $1-\delta$, in a polynomial in $(N,1/\varepsilon,1/\delta)$ time.

Using   the Gaussian approximation and the boson birthday paradox we show below that, under a  condition on the mode mismatch, the NDBS does exactly what is asked in   the \textit{$|BS|^2_\pm$-problem}, i.e. what the classical computer cannot do. We employ  Chebyshev's probability inequality \cite{Gnedenko}, stating that for  a random variable $X$ with $\langle X\rangle =0$,  the probability $\mathcal{P}\left({|X|}/{\sqrt{\langle X^2\rangle}}\ge 1/s\right)\le s^2$, for any $s>0$. Using that the $U_{kl}$ are i.i.d. random variables with the probability   density (\ref{EQ4}),  that  $J(I) = 1$  ($I$ is the  identity permutation), and $\langle U_{kl}\rangle = 0$ we obtain from Eqs. (\ref{EQ2})-(\ref{EQ3})
\begin{eqnarray}
\label{EQ6}
&& \langle P_0 \!- \!P_\eta\rangle  = \sum_{\sigma_1,\sigma_2}[1 \!-\! J(\sigma_{21})]\langle\prod_{\alpha=1}^N U^*_{k_{\sigma_1(\alpha)},l_\alpha} U_{k_{\sigma_2(\alpha)},l_\alpha}\rangle  \nonumber\\
&& =  \sum_{\sigma_1,\sigma_2}[1 \!-\! J(\sigma_{21})]\delta_{\sigma_1,\sigma_2}\prod_{\alpha=1}^N \langle|U_{k_{\sigma_1(\alpha)},l_\alpha}|^2\rangle  =0.
\end{eqnarray}
Similarly,  after more involved calculations (see Appendix \ref{secA3}), we get 
\begin{eqnarray}
\label{EQ7}
&& \!\!\langle(P_0-P_\eta)^2\rangle = \sum_{\sigma^{}_1,\sigma^{}_2}\sum_{\sigma^\prime_1,\sigma^\prime_2}[1-J(\sigma_{21})][1-J(\sigma_{21}^\prime)]\nonumber\\
&& \times\langle\prod_{\alpha=1}^N U^*_{k_{\sigma_1(\alpha)},l_\alpha} U_{k_{\sigma_2(\alpha)},l_\alpha}
U^*_{k_{\sigma_1^\prime(\alpha)},l_\alpha} U_{k_{\sigma_2^\prime(\alpha)},l_\alpha}\rangle\nonumber\\
&&= \left(\frac{N!}{M^N}\right)^2\frac{1}{N!}\sum_{\sigma}\chi(C_1(\sigma))\left[1-J(\sigma)\right]^2,
\end{eqnarray}
where we have defined  $\chi(n) = n!\sum_{k=0}^n\frac{1}{k!} =  \int\limits_1^\infty \rd z\, z^n e^{1-z}$.  
Let us introduce a  rescaled  variance  
\be
\mathcal{V}(N,\eta) =  \frac{1}{N!}\sum_{\sigma}\chi(C_1(\sigma))\left[1-J(\sigma)\right]^2.
\en{V}
Now, the inequality complementary to  Chebyshev's one reads  (for $\varepsilon>0$)
\be
\mathcal{P}\left(|P_0-P_\eta |<\varepsilon\frac{N!}{M^N}\right) >1- \frac{\mathcal{V}(N,\eta)}{\varepsilon^2},
\en{EQ8}
where    Eqs. (\ref{EQ7}) and (\ref{V}) were used. Eq. (\ref{EQ8}) resembles  the statement of the \textit{$|BS|^2_\pm$-problem}: if we are able to control the cycle sum  $\mathcal{V}(N,\eta)$, i.e. by varying the classicality parameter $\eta$, such that the r.h.s. in Eq. (\ref{EQ8}) stays close to $1$ then  the NDBS, with success probability close to $1$,  approximates the ideal BS of Aaronson and Arkhipov to within an additive  error (in the  required form).   Let us now formalize this statement. \textit{Given  an error $\varepsilon$ and a success probability $1-\delta$, if  the rescaled variance  $\mathcal{V}(N,\eta)$ (\ref{V}) observes  the bound   }
\be
\mathcal{V}(N,\eta) \le \varepsilon^2{\delta},
\en{EQ12}
\textit{then the NDBS solves  the $|BS|^2_\pm$-problem, i.e. performs a computational  task   which cannot be simulated  on a classical computer with only polynomial resources.} Eq. (\ref{EQ12}) is a sufficient bound which may be not necessary for the NDBS to outperform the classical computers,  since  Chebyshev's inequality can be a crude  approximation. However,  it   usually  captures  the scaling of  the tail probability of  a random variable in terms of its variance.   Eq.~(\ref{EQ12})  states that   the $N$-scaling   of  the minimal approximation error with which  the NDBS  satisfies the  \textit{$|BS|^2_\pm$-problem} is defined by  the rescaled variance  $\mathcal{V}(N,\eta)$.

\begin{figure}[htb]
\begin{center}
\includegraphics[width=0.75\textwidth]{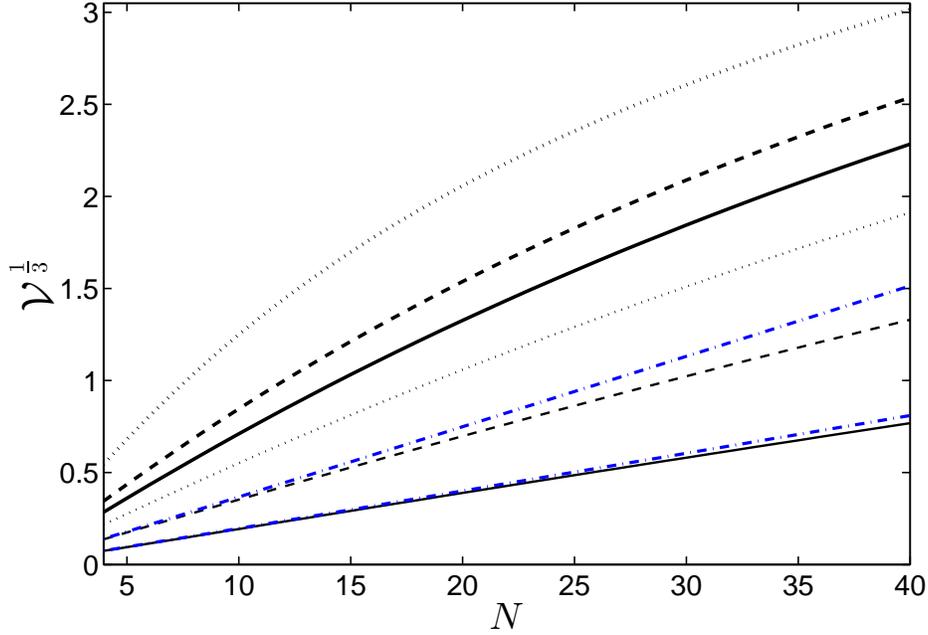}
\caption{ (Color online) Behavior of  cubic root of the  reduced variance $ \mathcal{V}(N,\eta)$   for several values of the two-photon indistinguishability parameter $g_2$ (from bottom to top): $g_2=0.99$ (thin solid line), $g_2=0.975$ (thin dashed line), $g_2 = 0.95$ (thin dotted line), $g_2 =0.925$ (thick solid  line), $g_2 =0.9$ (thick dashed line), and $g_2=0.8$ (thick dotted line). We have used the Gaussian model of the single photons with  random arrival times. The two dash-dotted lines give the approximation following from  Eq.~(\ref{EQ9}).    }\label{Fig2}
\end{center}
\end{figure}

Eq.~(\ref{EQ12}) involves the cycle sum (\ref{V}) computable only numerically  for each particular density matrix $\rho$ depending on the  sources. Let us analyze in detail   the model of   single photons with random arrival times, discussed above,   taking  both $\Phi(\tau,\omega)=\phi(\omega)e^{i\omega\tau}$ and $p(\tau)$ to be  Gaussian distributions, e.g. spectrally-shaped  by the stimulated Raman  technique of Ref. \cite{STIRAP}  with the Gaussian distributed random arrival times (centered at $\tau=0$):
\begin{eqnarray}
&& \Phi(\tau,\omega) = \frac{1}{\sqrt{2\pi}\Delta\omega}\exp\left(i\omega\tau-\frac{(\omega-\omega_0)^2}{2\Delta\omega^2}\right),
\\
 && p(\tau) =  \frac{1}{\sqrt{2\pi}\Delta\tau}\exp\left( -\frac{\tau^2}{2\Delta\tau^2}\right).
\end{eqnarray}
In this case, all integrals in Eq. (\ref{gk}) are Gaussian and can be evaluated. Such a model   also is interesting from the point of view of practical optimality, since as shown in  Ref. \cite{GAUSS}, the Gaussian  shaped form of single photons is optimal for interference experiments.   
Setting $\gamma = \frac{2\eta^2}{1+2\eta^2}$, we obtain $g_k$ as  a positive monotonously decreasing  function of $\gamma$ (and, hence, of $\eta^2$):
\be
g_k = (1- \gamma)^\frac{k}{2}(1-\gamma^k)^{-\frac{1}{2}}.
\en{G1}
An elementary algebra gives 
\be
J(\sigma) =  (1- \gamma)^\frac{N}{2}\prod_{k=1}^N (1-\gamma^k)^{-\frac{C_k(\sigma)}{2}}.
\en{G3}
In this case, one can also express $g_k$ and, hence, $J$ as functions of $g_2$ only, since $g^2_2$ and $\gamma$ are   M\"obius transformations of each other.  We have $\gamma = (1-g^2_2)/(1+g^2_2)$  (and  $\eta^2 = (g_2^{-2}-1)/2$). 
Moreover, $g_2 = \langle\mathcal{F}\rangle/\sqrt{2- \langle\mathcal{F}\rangle^2}$. For this model, the results are presented in Fig.~\ref{Fig2}, where we plot the cubic root of $\mathcal{V}(N,\eta)$.

For  a small two-photon distinguishability  $1-g_2\approx 2\eta^2\ll 1$ (i.e. for a small mode mismatch), the dependence of $\mathcal{V}^\frac{1}{3}(N,\eta)$ on $N$ in Fig.~\ref{Fig2} is approximately  a linear function. This is a general feature. Indeed,  as  shown above,   $g_k(\eta) \approx 1 - k\eta^2 $ for $\eta\ll1$  and    $J(\sigma)\approx 1-\eta^2[N-C_1(\sigma)]^2$. Inserting this into the definition of $\mathcal{V}(N,\eta)$   and taking the integral over $z$  in  the resulting expression (coming from the integral representation of  $\chi(C_1)$ in Eq. (\ref{V})) we get after an elementary algebra
 \be
\mathcal{V}(N,\eta) \approx \eta^4 \left(\frac{N^3}{3} - \frac{N^2}{2} +\frac{7N}{6} - 1\right).
\en {EQ9}
Eq. (\ref{EQ9})  for $N\gg1$   reveals the scaling
$\mathcal{V}(N,\eta) \approx \eta^4 N^3/3 \approx (1-\langle \mathcal{F}\rangle )^2N^3/3 $.
Therefore, the photon mode mismatch  ($1-\langle \mathcal{F}\rangle$) must scale  approximately as $N^{-3/2}$  in the total number of photons, if  the NDBS is to be scaled up while keeping the product     $\varepsilon^2\delta$ constant (i.e. at the same level of practical hardness of classical simulation). As seen from Fig.~\ref{Fig2}, the approximation (\ref{EQ9}) deviates from  the exact result  for sufficiently  large $N$, where the contribution from the higher-order terms $\sim \eta^p$, $p>2$, becomes important. Such higher-order   terms  are model specific and thus cannot be obtained in the general form. The optimality of the Gaussian model suggest that Fig. \ref{Fig2} shows the optimal instance of the bound (\ref{EQ12}).  

 Our main  result (\ref{EQ12})    provides also a sufficient condition for approximation of the BS  by the  NDBS in the variational distance, i.e. as in Eq. (\ref{VarDist}), but for a fraction $1 -\frac{\mathcal{V}(N,\eta)}{4\varepsilon^2}$ of the network matrices $U$. Indeed,  the 1-norm (known in the probability theory as the variational distance) is bounded as  $|| \mathcal{D}_0-\mathcal{D}_\eta||^2\le \frac{1}{4}(\sum_{\vec{l}}1)\sum_{\vec{l}} [P_0(\vec{l}|\vec{k}) - P_\eta(\vec{l}|\vec{k})]^2$. Using this upper bound and applying Chebyshev's inequality to  $ ||\mathcal{D}_0-\mathcal{D}_\eta||^\prime $  of Eq. (\ref{VarDist}) considered as  random variable on the Haar measure, we get
\be
\mathcal{P}\left( ||\mathcal{D}_0-\mathcal{D}_\eta||^\prime < \varepsilon \right) > 1- \frac{\mathcal{V}(N,\eta)}{4\varepsilon^2}.
\en{EQ14}

The experimental demonstration  of the NDBS  operation beyond the power of  classical  computers  could proceed in showing that, for  a randomly chosen   network matrix, the  NDBS with a  fixed mode mismatch approximates the output probabilities of the  ideal BS of Aaronson and Arkhipov  to within an error $\pm \varepsilon \frac{N!}{M^N}$, i.e. solves the computational task specified in the $|BS|^2_\pm$-problem,  where  the product of the squared error $\varepsilon^2$ and the failure probability $\delta$  (i.e. the Haar measure of the excluded network matrices)   is at least as the reduced variance $\mathcal{V}(N,\eta)$.    The probabilities of the  ideal BS computer can still be obtained for $N\sim 20$  by   numerical  simulations.

\section{Conclusion}
\label{sec4}

In conclusion, we have considered a nondeterministic model of the BS computer, the NDBS, which generalizes the ideal BS computer of Aaronson  and Arkhipov \cite{AA} and captures the essential features of a realistic BS device with only partially indistinguishable single photons at the input.   If the average mutual fidelity of the   single photons satisfies  the derived   $N$-dependent bound,  the NDBS  device   cannot be efficiently  simulated on a  classical computer.  The sufficient condition derived in this work may be not necessary for the NDBS   to be hard to simulate classically, however, it  reveals the inverse $3/2$-power law scaling of the photon mode mismatch on the total number of photons   for scalability of the NDBS computer  at the same level of  practical hardness of its classical simulation (i.e. for the constant approximation error and fixed success probability with which the NDBS approximates the ideal BS in the  variational distance). Moreover, the results are also applicable to any other realization of the BS with identical single photon sources, for instance, with the Gaussian input states, proposed recently in Ref. \cite{GaussBS}, where the imperfect  indistinguishability of the   heralded  single photons  can be treated in similar way.

We have studied the so-called ``dilute limit" of unitary  $M$-mode network with $N$ bosons, i.e. with  $M\gg N^2$, for which the classical hardness is established, and when the average probability (over the random network matrices in the  Haar measure)  of two bosons landing at the same output mode is vanishing as $\mathcal{O}(\frac{N^2}{M})$. One might wonder, why then  the output probability distribution  of bosons is exponentially  harder to compute than that  of  fermions in a similar setup? Since this  question belongs to  the field of computational complexity theory, the  answer must be formulated in its  terms:    bosonic amplitudes are given by matrix permanents, while fermonic ones by matrix determinants, where the permanent  requires an  exponential in $N$  computation time, whereas the determinant is known to be polynomial in $N$. 

However, a physicist can be left unsatisfied by the permanent vs. determinant explanation, though absolutely correct, and try  inquire further:  what  specific feature of the bosonic  statistics  could be held responsible for this drastic difference, especially in view that the output rarely contains two bosons at the same mode?  One plausible candidate  is   the very same  bosonic bunching,  which is   unimportant at the output, but not \textit{during} the propagation in the network. Indeed, let us compare bosonic  and fermionic propagation through a unitary network,   bringing the two cases  to a ``common ground" by  decomposing  the unitary map between the input and output Fock states into a product of infinitesimal unitary  maps, i.e. using a Feynman type   sum over  the  paths, but now in the Fock space.  Such an  expansion involves summation over all intermediate occupation numbers and each term is a product of permanents (bosons) or determinants (fermions). In both cases, each factor in the product of amplitudes  becomes  easily computable for an infinitesimal unitary map   (to a sufficient approximation)  when the number of factors becomes sufficiently large.  But, as soon  as the number of infinitesimal maps in the product  grows  above the ratio $M/N^2$ it would be necessary to sum over the multiple occupation numbers for  bosons, i.e. bosonic   bunching would contribute in the  intermediate Fock states, whereas in  fermionic  case the occupation numbers would remain bounded by 1.  In the limit, when the Feynman type  expansion becomes exact,  one   recovers full bosonic  bunching as allowed by their statistics, while at the output it is still  negligible. Therefore, reformulating slightly Wigderson's famous joke  \cite{AA}, we can conclude  saying that  to arrive  at the same output configuration   as fermions, bosons have a much   harder job indeed, since they must go along a much larger  set of paths.

\acknowledgements
This work was supported by   the  CNPq   of Brazil. The author is indebted to  Scott Aaronson for  helpful  comments. 

\appendix

\section{Derivation of the probability formula}
\label{secA1}

We consider the case of single  photons which are   emitted by identical photon sources and   launched into distinct  modes $k_1, ..., k_N$   of a $M$-mode   linear optical  network with the  unitary matrix $U$  relating  the  input $a_k(\omega)$ and output $b_l(\omega)$   modes of frequency $\omega$, $a^\dag_k (\omega)= \sum_{l=1}^M U_{kl}b^\dag_l(\omega)$.   The input  state originated   from  a set of $N$ independent identical sources  of single photons is   given by the density matrix\begin{eqnarray}
\label{rhoN}
&&\rho^{(in)} = \int\rd \mathbf{x}_1 \ldots \int\rd\mathbf{x}_N \prod_{\alpha=1}^Np(\mathbf{x}_\alpha)\nonumber\\
&& \times |\Psi(\mathbf{x}_1,...,\mathbf{x}_N)\rangle \langle \Psi(\mathbf{x}_1,...,\mathbf{x}_N)|,
\end{eqnarray}
 where the Fock state $ |\Psi(\mathbf{x}_1,...,\mathbf{x}_N)\rangle$ is given in Eq. (\ref{EQ1}) of section \ref{sec2}.   The probability  of detecting $m_1,...,m_M$ photons in the output modes   described by the annihilation operators $b_1(t),...,b_M(t)$  can be derived by the standard quantum photon counting  theory \cite{Mandel,Glauber,KK,VWBook}. It is  in the form of an average on the density matrix (\ref{rhoN})
\be
p_{m_1,..,m_M} = \frac{1}{\prod_{l=1}^Mm_l!}\langle : \prod_{l=1}^M\mathcal{I}^{m_l}_l \exp\{-\sum_{l=1}^M \mathcal{I}_l\}: \rangle,
\en{E1}
where the double dots denote the time and normal ordering of the creation and annihilation operators and the detection operator reads 
\be
\mathcal{I}_l = \int\limits_t^{t+\Delta t}\rd \tau \int\limits_t^{t+\Delta t}\rd \tau^\prime G(\tau-\tau^\prime)b^\dag_l(\tau)b_l(\tau^\prime)
\en{E2}
with the detector efficiency  described by the  function $G(t)$. In our case, the initial state is a Fock state of $N$ single  photons in distinct modes and we postselect on the cases when  all $N$ photons are detected, $\sum m_l = N$. In this case the exponent in Eq. (\ref{E1}) does not contribute. Substituting  the Fourier expansions
\be
b_l(t) = \int\limits_0^{\infty}\frac{\rd\omega}{\sqrt{2\pi}}e^{-i\omega t} b_l(\omega)
\en{E3}
and (see, for instance, Ref. \cite{VWBook})
\be
G(t) =  \int\limits_0^{\infty}\frac{\rd\omega}{2\pi}e^{-i\omega t} \Gamma(\omega), \quad \Gamma(\omega)>0, 
\en{E4}
in Eq. (\ref{E1}), inserting the projector onto the vacuum $|0\rangle\langle0|$ between the creation and annihilation operators (since all photons are detected this changes nothing) and integrating over the times we obtain that the probability is given by the average of  the following operator
\begin{eqnarray}
\label{E5}
&&\Pi(m_1,...,m_M) = \frac{1}{\prod_{l=1}^M m_l!}\int\limits_0^{\infty} \rd \omega_1\cdot\ldots\cdot\int\limits_0^{\infty} \rd \omega_N \prod_{l=1}^N\Gamma(\omega_l)\nonumber\\
&&\times\left[ \prod_{\alpha=1}^N b^\dag_{l_\alpha}(\omega_\alpha)\right]|0\rangle\langle 0|\left[\prod_{\alpha=1}^N b_{l_\alpha}(\omega_\alpha)\right],
\end{eqnarray}
where the combined  index $(l_1,...,l_N)$ (the order being insignificant)   is the set $\{1,...,1,2,..,2,...,M,...,M\}$ with index $k$ appearing $m_k$ times. Thus, the output probability  of detecting $m_1,...,m_M$ photons in modes $1,...,M$  becomes
\begin{eqnarray}
\label{E7}
&& P(m_1,...,m_M|k_1,...,k_N) =  \int\rd\mathbf{x}_1\!\cdot\!\ldots\!\cdot\!\int\rd\mathbf{x}_N\prod_{\alpha=1}^Np(\mathbf{x}_\alpha)  \nonumber\\
&& \times\langle \Psi(\mathbf{x}_1,...,\mathbf{x}_N)|\Pi(m_1,...,m_M)|\Psi(\mathbf{x}_1,...,\mathbf{x}_N)\rangle.
\end{eqnarray} 
 The  operators $\Pi(m_1,\ldots,m_M)$ are positive Hermitian, but they generally do not sum up to the identity operator (more precisely, to the projector on the symmetric subspace of $N$ bosons) for $m_1+\ldots+m_M = N$.  However, for efficient  detectors, when  all output  photons are detected, after a suitable normalization (see below)  $\Pi(m_1,\ldots,m_M)$ become     the POVM elements realizing  the above described detection. In this case the probabilities in Eq. (\ref{E7}) sum to $1$ under the constraint $m_1+\ldots+m_M = N$. 
Using the evolution in the unitary network 
\be
a^\dag_k(\omega) = \sum_{l=1}^M U_{kl} b^\dag_l(\omega)
\en{E8}
and the identity 
\begin{eqnarray}
\label{E9}
&&\langle0|\left[\prod_{\alpha=1}^Nb_{l^\prime_\alpha}(\omega^\prime_\alpha)\right]\left[\prod_{\alpha=1}^N b^\dag_{l_\alpha}(\omega_\alpha)\right]|0\rangle\nonumber \\
&& =\sum_\sigma \delta_{l^\prime_\alpha,l_{\sigma^{-1}(\alpha)}}\delta(\omega^\prime_\alpha - \omega_{\sigma^{-1}(\alpha)}),
\end{eqnarray}
where $\sigma$ is a permutation, we obtain (transferring the permutations $\sigma_{1,2}$  from the two inner products, as in Eq. (\ref{E9}), to the $k$-indices)
\begin{eqnarray}
\label{E10}
&& P(m_1,...,m_M|k_1,...,k_N) = \frac{1}{\prod_{l=1}^M m_l!}\sum_{\sigma_1}\sum_{\sigma_2} J(\sigma_2\sigma^{-1}_1) \nonumber\\
&& \quad \times \prod_{\alpha=1}^N U^*_{k_{\sigma_1(\alpha)},l_\alpha} U_{k_{\sigma_2(\alpha)},l_\alpha},
\end{eqnarray} 
with $J$  given  as follows (the product is a shortcut notation for the multiple integration over $\omega_\alpha$ and $\mathbf{x}_\alpha$)
\begin{eqnarray}
\label{E11}
&&J(\sigma) = \prod_{\alpha=1}^N\int\rd\mathbf{x}_\alpha p(\mathbf{x}_\alpha)\int\limits_0^\infty\rd\omega_\alpha \Gamma(\omega_\alpha)\nonumber\\
&&\times\phi^*(\mathbf{x}_\alpha,\omega_\alpha)\phi(\mathbf{x}_\alpha,\omega_{\sigma^{-1}(\alpha)}).
\end{eqnarray}
Here we have used the symmetry of the multiple  integral under permutation of the integration  variables, reassigning the variables as $\omega_\alpha \equiv \omega_{\sigma^{-1}_1(\alpha)} $ and defining  $\sigma \equiv \sigma_2\sigma^{-1}_1$.

The structure of the integrals in Eq. (\ref{E11}) makes  $J$ factorize into a product of similar functions depending on the cycles from the cycle decomposition of the  permutation $\sigma$ (since each of the two multiple integrals, one over $\omega_\alpha$ and one over  $\mathbf{x}_\alpha$, factorizes). Moreover,  by the  above mentioned  permutational symmetry of the integration variables,   the cycles with the same number of elements contribute the same  factor. Therefore we obtain 
\begin{eqnarray}
\label{E12}
&&J(\sigma)  = \prod_{k=2}^N g_k^{C_k(\sigma)},\\
&& g_k \equiv  \prod_{\alpha=1}^k\int\limits_0^\infty\rd \omega_\alpha 
\int\rd \mathbf{x}_\alpha p(\mathbf{x}_\alpha) \Phi(\mathbf{x}_\alpha,\omega_{\alpha-1})\Phi^*(\mathbf{x}_\alpha,\omega_\alpha),\nonumber\\
&&\label{E12A}\end{eqnarray}
 where the index $\alpha$ is cyclic ($\alpha=0$ is   $\alpha=k$),  $C_k$ is the number of cycles of length $k$,  with $\sum kC_k=N$ \cite{Stanley}, and  $\Phi(\mathbf{x},\omega) \equiv \sqrt{\Gamma(\omega)}\phi(\mathbf{x},\omega)$.


\section{Derivation of the expression for the variance of $P_0-P_\eta$}
\label{secA3}

We have  for the variance 
\begin{eqnarray}
\label{G4}
&& \langle(P_0-P_\eta)^2\rangle = \sum_{\sigma,\tilde{\sigma}}\sum_{\sigma_R,\tilde{\sigma}_R}[1-J(\tilde{\sigma}_R)][1-J(\sigma_R)]
\nonumber\\
&& \times\prod_{\alpha=1}^N \langle U^*_{k_{\sigma(\alpha)},l_\alpha} U_{k_{\sigma_R\sigma(\alpha)},l_\alpha}
U^*_{k_{\tilde{\sigma}(\alpha)},l_\alpha} U_{k_{\tilde{\sigma}_R\tilde{\sigma}(\alpha)},l_\alpha}\rangle\nonumber\\
\end{eqnarray}
where we have introduced  the relative  permutations $\sigma_R$ and $\tilde{\sigma}_R$ and taken into account the mutual independence of $U_{k_\beta,l_\alpha}$ for  the set of distinct indices $l_1,\ldots,l_N$.  The nonzero terms in the sum over all permutations in Eq. (\ref{G4}) occur  under the condition that for any  $\alpha \in \{1,\ldots,N\}$ either of the two sets of equations below is satisfied:
 \begin{eqnarray}
&& 
\sigma_R\sigma(\alpha)  = \sigma(\alpha),\quad \tilde{\sigma}_R\tilde{\sigma}(\alpha) = \tilde{\sigma}(\alpha),
\label{G5a}\\
&& 
\sigma_R\sigma(\alpha)  = \tilde{\sigma}(\alpha),\quad \tilde{\sigma}_R\tilde{\sigma}(\alpha) = \sigma(\alpha).
\label{G5b}
\end{eqnarray}
For each choice   of the permutations $\{\sigma,\tilde{\sigma},\sigma_R,\tilde{\sigma}_R\}$ denote the ordered (in some way) set of  all $\alpha$ satisfying Eq. (\ref{G5a}) as $\alpha^{(I)}$ and the  ordered set of the rest of the indices as $\alpha^{(II)}$ (these satisfy Eq. (\ref{G5b})) (the two ordered sets give an ordered partition of the set of all indices $\{1,\ldots,N\}$). Introduce also the ordered sets $\beta^{(I)}$ and $\beta^{(II)}$ and their versions with the tilde,  $\tilde{\beta}^{(I)}$ and $\tilde{\beta}^{(II)}$, as the result of action of $\sigma$ (respectfully, $\tilde{\sigma}$) on the sets $\alpha^{(I)}$ and $\alpha^{(II)}$, i.e. by $\beta_j = \sigma(\alpha_j)$ and $\tilde{\beta}_j = \tilde{\sigma}(\alpha_j)$. Each $\beta$-set and its version with the tilde are permutations of each other:  $\tilde{\beta}^{(I,II)}_j  = \tilde{\sigma}\sigma^{-1}(\beta^{(I,II)}_j)$. 
Eq. (\ref{G5a}) states that $\beta^{(I)}_j$ and $\tilde{\beta}^{(I)}_j$, $j=1,\ldots, |\alpha^{(I)}|$,  are   fix points (i.e. $1$-cycles) of the permutations $\sigma_R$ and $\tilde{\sigma}_R$, respectfully (thus the sets of their  fix points  coincide). Eq. (\ref{G5b}) states that $\tilde{\sigma}_R$  is inverse to  $\sigma_R$ acting on $\beta^{(II)}$, i.e. $\sigma_R(\beta^{(II)}_j) = \tilde{\beta}^{(II)}_j$ and   $\tilde{\sigma}_R(\tilde{\beta}^{(II)}_j) = \beta^{(II)}_j$, $j=1,\ldots, |\alpha^{(II)}|$. From these facts  the necessary conditions for nonzero contribution in Eq. (\ref{G4})  follow:
\be
\tilde{\sigma}_R = \sigma^{-1}_R,\quad \tilde{\sigma} = (\tau_1\otimes I_2) \sigma_R\sigma,
\en{G6}
where $\tau_1$ is an   \textit{arbitrary} permutation of the set $\beta^{(I)}$ and $I_2$ is the identity permutation of the set $\beta^{(II)}$. Note also that the number of all indices $\alpha^{(I)}$ satisfies  $|\alpha^{(I)}| = C_1(\sigma_R)$, where $C_1$ is the number of 1-cycles (fix points) of the permutation. Let us now use Eqs. (\ref{G5a}), (\ref{G5b}), and (\ref{G6}) into Eq. (\ref{G4}). Under the Gaussian approximation in   Eq. (\ref{EQ4}) of section \ref{sec3} $\langle |U_{kl}|^2\rangle = 1/M$ and $\langle |U_{kl}|^4\rangle = 2/M^2$. Hence, we obtain for $\alpha\in \alpha^{(I)}$:
\begin{eqnarray}
\label{G7}
&&\prod_{\alpha\in\alpha^{(I)}} \langle |U_{k\sigma(\alpha),l_\alpha}|^2 |U_{k\tilde{\sigma}(\alpha),l_\alpha}|^2\rangle \nonumber\\
&&= \left(\frac{1}{M}\right)^{2[|\alpha^{(I)}|-C_1(\tilde{\sigma}\sigma^{-1})]} \left(\frac{2}{M^2}\right)^{C_1(\tilde{\sigma}\sigma^{-1})} \nonumber\\
&&= 2^{C_1(\tau_1)} \left(\frac{1}{M}\right)^{2C_1(\sigma_R)},
\end{eqnarray}
where we have taken into account  that, since all fix point of $\sigma_R$ are in $\beta^{(I)}$,  all fix points of $\tilde{\sigma}\sigma^{-1}$ belong to the set $\beta^{(I)}$ and are also fix points of $\tau_1$. Hence,  using that $|\alpha^{(II)}|=N-|\alpha^{(I)}|= N-C_1(\sigma_R)$, for $\alpha\in \alpha^{(II)}$ we obtain 
\be
\prod_{\alpha\in\alpha^{(II)}} \langle |U_{k\sigma(\alpha),l_\alpha}|^2 |U_{k\tilde{\sigma}(\alpha),l_\alpha}|^2\rangle =   \left(\frac{1}{M}\right)^{2[N-C_1(\sigma_R)]}.
\en{G8}
Inserting the results of  Eqs. (\ref{G7}) and (\ref{G8})  into Eq. (\ref{G4}), performing the summation over the  independent (free) permutations $\sigma$, $\sigma_R$, and $\tau_1$, and using that $J(\sigma^{-1}) = J(\sigma)$ (since the  inverse permutation  has the same cycle structure) we obtain the following  expression for the variance 
\be
 \langle(P_0-P_\eta)^2\rangle = \left(\frac{N!}{M^N}\right)^2\frac{1}{N!}\sum_{\sigma_R} \chi(C_1(\sigma_R))\left[1-J(\sigma_R)\right]^2.
\en{G9}
Here $\chi(n)$ is the  cycle sum 
\be
\chi(n) \equiv \sum_{\tau }2^{C_1(\tau)}  = n!\sum_{k=0}^n\frac{1}{k!} = \int\limits_1^\infty \rd t\, t^n e^{1-t},
\en{G10}
where $\tau$ is a permutation   of $n$ elements (see, for instance, Ref. \cite{Stanley}).


\section{The cycle sum $\mathcal{V}(N,\eta)$   }
\label{secA4}

One can express the summation over the permutations in the definition of $\mathcal{V}$ (note that there are, in total,  $N!$ terms),
\be
\mathcal{V}(N,\eta)= \frac{1}{N!}\sum_{\sigma}\chi(C_1(\sigma))\left[1- \prod_{k=2}^Ng^{C_k(\sigma)}_k(\eta) \right]^2,
\en{D1}
as  summation over all  partitions of $N$ into a sum of positive  integers.  Indeed, there are 
$N!/(\prod_{k=1}^N k^{C_k} C_k!)$ permutations with the cycle structure $(C_1,\ldots,C_N)$ (see, for instance, Ref. \cite{Stanley}), the summation is over the integer partitions of $N$ into the sum of integers,   from $1$ to $N$, where each integer $k$ corresponds to  a  cycle length in the cycle structure of the permutation, while  the multiplicity is $C_k$. We get
\be
\mathcal{V} = \sum_{(C_1,\ldots,C_N)}  \frac{\chi(C_1)\left(1 - \prod_{k=2}^Ng^{C_k}_k \right)^2}{\prod_{k=1}^Nk^{C_k}C_k!},
\en{D2}
where the summation is under the constraint that $ \quad \sum_{k=1}^NkC_k=N$. The sum in Eq. (\ref{D2}) can  be efficiently  calculated   numerically,  if  $N$ is not very large. 



\begin{thebibliography}{99}


\bibitem{AA} S. Aaronson and A. Arkhipov,  Theory of Computing \textbf{9},  143 (2013).


\bibitem{SPS} See, for instance, the reviews: B. Lounis and M. Orrit, Rep. Prog. Phys. \textbf{68},  1129 (2005); G. S. Buller and R. J. Collins,   Meas. Sci. Technol. \textbf{21}, 012002 (2010); M. D. Eisaman, J. Fan, A. Migdall, and S. V. Polyakov,  Rev.  Sci. Instr. \textbf{82}, 071101 (2011).

\bibitem{HOM} C. K. Hong, Z. Y. Ou, and L. Mandel, Phys. Rev. Lett. \textbf{59} (1987)  2044.

\bibitem{LB}  Y. L. Lim and A. Beige, New J. Phys., \textbf{7}  155  (2005).

\bibitem{MPI} M. C. Tichy  \textit{et al},  New J.  Phys., \textbf{14}  (2012) 093015.

\bibitem{Shor} P. Shor, in \textit{Proceedings of the 35th Annual Symposium on Foundations of Computer Science} (IEEE Comp. Soc. Press, Los Alamos, CA, 1994), p. 124; SIAM J. Comput. \textbf{26}, 1484 (1997).



\bibitem{BookNC} M. A. Nielsen and I. L. Chuang, \textit{Quantum Computation and Quantum Information} (Cambridge Univ. Press, 2000).

\bibitem{Minc} H. Minc, \textit{Permanents, Encyclopedia of Mathematics and Its Applications}, Vol. \textbf{6} (Addison-Wesley Publ. Co., Reading, Mass., 1978).

\bibitem{C} E. R. Caianiello, Nuovo Cimento, \textbf{10}, 1634 (1953); \textit{Combinatorics and Renormalization in Quantum Field Theory}, Frontiers in Physics, Lecture Note Series (W. A. Benjamin, Reading, MA, 1973).

\bibitem{S} S. Scheel, \textit{Permanents in linear optical networks}, quant-ph/0406127 (2004).


\bibitem{Valiant} L. G. Valiant, Theoretical Coput. Sci., \textbf{8},  189  (1979).

\bibitem{A1} S. Aaronson, Proc. Roy. Soc. London A, \textbf{467},  (2008) 3393.

\bibitem{Ryser} H. Ryser, \textit{Combinatorial Mathematics}, Carus Mathematical Monograph No. 14. (Wiley, 1963).


\bibitem{E1} M. A. Broome \textit{et al},  Science \textbf{339},   794 (2013).

\bibitem{E2} J. B. Spring \textit{et al}, Science, \textbf{339},  798 (2013).

\bibitem{E3}  M. Tillmann \textit{et al},  Nature Photonics,  \textbf{7}, 540 (2013).

\bibitem{E4} A. Crespi \textit{et al},   Nature Photonics, \textbf{7}, 545 (2013).

\bibitem{ErrCorr}  D. Aharonov and M. Ben-Or, In Proc. ACM STOC p. 176 (1997); quant-ph/9703054.

\bibitem{ErrCorrRev} E. Knill, R. Laflamme, and W. Zurek, Science \textbf{279}, 342 (1998).

\bibitem{Steane} A. M. Steane, Phys. Rev. A  \textbf{68}, 042322 (2003).

\bibitem{Knill} E. Knill, Nature \textbf{434}, 39 (2005).



\bibitem{RR} P. P. Rohde and T. C. Ralph, Phys. Rev. A \textbf{85}, 022332  (2012).

\bibitem{R1} P. P. Rohde, Phys. Rev. A \textbf{86}, 052321 (2012).

\bibitem{NoisyBS} A. Leverrier and R. Garc{\'i}a-Patr{\'o}n, \textit{Does Boson Sampling need Fault-Tolerance?}, quant-ph/1309.4687 (2013). 


\bibitem{JSV} M. Jerrum, A. Sinclair, and E. Vigoda, Journal of the ACM \textbf{51},  671 (2004).



\bibitem{Multiplex} M. J. Collins \textit{et al}, Nature Comm. \textbf{4}, 2582 (2013).


\bibitem{NDSPS1}  E. B. Flagg \textit{et al}, Phys. Rev. Lett. \textbf{104}, 137401 (2010).

\bibitem{NDSPS2} R. B. Patel \textit{et al}, Nature Photonics, \textbf{4}, 632 (2010).

\bibitem{NDSPS3}  Y.-M. He \textit{et al},     Nat. Nanotech. \textbf{8},  213 (2013).

 

\bibitem{Mandel} L. Mandel, in \textit{ Progress in Optics},  ed. by E. Wolf (North-
Holland, Amsterdam, 1963), Vol. 2, p. 181.

\bibitem{Glauber}  R. J. Glauber,   \textit{ Optical Coherence and Photon Statistics}, (Gordon
\& Breach, New York, 1965), p. 65.

\bibitem{KK} P. L. Kelley and W. H. Kleiner, Phys. Rev. \textbf{136}, A316 (1964).




\bibitem{Stanley}   R. P. Stanley, \textit{Enumerative Combinatorics}, 2nd ed., Vol. 1 (Cambridge University Press, 2011).

\bibitem{Young} R. A. Campos, B. E. A. Saleh, and M. C. Teich, Phys. Rev. A \textbf{40}, 1371 (1989);
S.-H. Tan, Y. Y.  Gao, H. de Guise, and B. C. Sanders, Phys. Rev. Lett. \textbf{110}, 113603 (2013).


\bibitem{Ou}  Z. Y. Ou, Phys. Rev. A \textbf{74}, 063808 (2006).

\bibitem{RMS} P. P. Rohde, W. Mauerer and C. Silberhorn, New J. Phys. \textbf{9}, 91 (2007).



\bibitem{AG} A. Arkhipov and G.Kuperberg, Geom. Topol. Monogr., \textbf{18}, 1 (2012).

\bibitem{BBPExp} N. Spagnolo \textit{et al},  Phys. Rev. Lett. \textbf{111}, 130503 (2013).



\bibitem{Gnedenko} B. V. Gnedenko,  \textit{The Theory of Probability} (English Translation; Mir Publishers, Moscow, 1978) p. 198.



\bibitem{STIRAP} A. Kuhn, M. Hennrich, and G. Rempe, Phys. Rev. Lett. \textbf{89}, 067901(2002). 



\bibitem{GAUSS} P. P. Rohde, T. C. Ralph,  and  M. A. Nielsen, Phys. Rev. A \textbf{72}, 052332 (2005).


\bibitem{GaussBS} A. P. Lund \textit{et al}, \textit{Boson Sampling from Gaussian States},  quant-ph/1305.4346 (2013). 

\bibitem{VWBook} W. Vogel and D. G. Welsch, \textit{Quantum Optics},  (Wiley-VCH Verlag GmbH \& Co. KGaA, 2006), p. 173. 



\end{thebibliography}
\end{document}